\documentclass[12pt]{article}

\begin{document}

\begin{center}

{\Large {"Old" Conformal Bootstrap in the AdS/CFT Context}}

\vspace{1,5cm}

{Boris L. Altshuler}\footnote{E-mail addresses: baltshuler@yandex.ru $\,\,\,  \& \,\,\,$  altshul@lpi.ru}

\vspace{1cm}

{\it Theoretical Physics Department, P.N. Lebedev Physical
Institute, \\  53 Leninsky Prospect, Moscow, 119991, Russia}

\vspace{1,5cm}

\end{center}

{\bf Abstract:} The Schwinger-Dyson equations for Green functions and 3-vertexes in planar approximation used 50 years ago in the earlier theory of conformal bootstrap are written in frames of the AdS/CFT correspondence. The technique 
of "Wightman-Witten" diagrams is used in calculations of the bubble bulk diagrams that permitted to obtain the values of conformal dimensions of primary operators in two models of the interacting bulk scalar fields which conformal dimensions obey "extremal" relation. Simple expressions for the four-point tree "Wightman-Witten" correlators of these primary operators are obtained and spectra of corresponding Bethe-Salpeter equations are calculated.

\vspace{0,5cm}


\newpage

\section{Introduction}

\quad This paper was inspired by the work \cite{Melon1} that Igor Klebanov has reported in the Lebedev Institute in 2017. In \cite{Melon1} (see also \cite{Melon2}), among other interesting things, the expressions for spectra of conformal dimensions in one-dimensional Sachdev-Ye-Kitaev and in $d$-dimensional field theory models were obtained from the field theory "bootstrap" equations for Green function $G(x_{1}, x_{2})$ and vertex $\Gamma (x_{1}, x_{2}, x_{3})$ expressed symbolically in most general form as:

\begin{eqnarray}
\label{1}
G (x_{1}, x_{2}) = \int\, G(x_{1}, x)\,\Sigma (x, y \, | \, G, \Gamma) \, G(y, x_{2})\, dx\,dy \, ;  \nonumber
\\
\\
\Gamma (x_{1}, x_{2}, x_{3}) = M^{\triangle} (x_{1}, x_{2}, x_{3} \, | \, G, \Gamma),  \qquad  \qquad \nonumber
\end{eqnarray}
where $\Sigma$ is quantum self-energy and $M^{\triangle}$ is a triangle 3-gamma diagram - both constructed from the same $G$ and $\Gamma$ that stand on the LHS in (\ref{1}), symbolically:

\begin{equation}
\label{2}
\Sigma = \Gamma \, G^{2} \, \Gamma; \,\,\,\,\, \, \, \, \, \, \,  M^{\triangle} = \Gamma\, G \, \Gamma \, G \, \Gamma \, G. 
\end{equation}
Bootstrap equation $\Gamma = M^{\triangle}$ in (1), that actually is a sort of Bethe-Salpeter equation, was considered in \cite{Melon1} in the limit $x_{3} \to \infty$, we also follow this way in Sec. 5. 

Eq-s (\ref{2}) mean that only planar diagrams are taken into account in (\ref{1}). Planar expansion is traditionally used in the self-consistent calculations of the anomalous conformal dimensions, beginning from the pioneer Migdal and Polyakov works on the "old" conformal bootstrap \cite{old1}, \cite{old2} which were developed in \cite{Parisi} - \cite{Dobrev2}, see for example \cite{Grensing} and references therein. In some models the choice of only planar diagrams may be justified as "most divergent" ones or in frames of the $1/N$ expansion. However in his 1982 Nobel Lecture \cite{Wilson} Kenneth Wilson criticized this approximation as ungrounded. Nevertheless people use it, and we shall do the same in the present paper, having in mind that interesting results are the best justification of any postulate.  

"Old" conformal bootstrap given by Eq-s (\ref{1}), (\ref{2}) should not be confused with most popular nowadays "OPE conformal bootstrap" based on the demand of crossing symmetry of the operator product expansions - see primary papers \cite{Ferrara} - \cite{Mack} and modern reviews \cite{Rychkov}, \cite{Alday}. This non-Lagrangian conformal bootstrap is in a sense close to the "nuclear democracy" bootstrap based upon the demand of $S$-matrix crossing symmetry \cite{Chew}.

Another background of the present paper are works \cite{Giombi1} where bubble Witten diagrams were calculated in case of triple bulk interactions, and \cite{Giombi2} where it is shown that Witten diagrams with "extremal" vertexes and bulk Green functions being replaced by the bulk Wightman functions are easily calculable, and that conformal IR divergences may be absorbed into "bare" bulk coupling constant giving the finite Witten diagrams expressed through the norm-invariant coupling constant.

Replacement in Witten diagrams of the bulk Green functions $G_{\Delta}$ with the homogeneous difference (Wightman function)

$$
{\widetilde G}_{\Delta} = G^{IR}_{\Delta} - G^{UV}_{d-\Delta}
$$ 
is one more technical postulate of the present paper permitting, following \cite{Giombi2}, to simplify the calculations. This technique was extensively used for different bulk fields in calculations of the double-trace flow of the quantum one-loop bulk vacuum energy when regular (IR) asymptotic boundary condition flows to irregular (UV) one  \cite{Mitra} - \cite{Alt2}. In most general form this means that instead of standard quantum functional $Z_{0} = [\rm{Det}G]^{-1/2}$ the ratio of two quantum functionals for two different boundary conditions of the corresponding Green functions $G$ is considered \cite{Barv}. We generalize this approach to the case of the non-zero source $j(X)$ of the bulk quantum field ($X$, $Y$ are bulk coordinates):

\begin{equation}
\label{3}
{\widetilde Z}_{free}(j(X)) = \frac{Z^{IR}_{free}(j)}{Z^{UV}_{free}(j)} = \frac{\sqrt{\rm{Det}^{UV}}}{\sqrt{\rm{Det}^{IR}}} \cdot e^{\int\int j(X)\,{\widetilde G}(X,Y)\,j(Y)\,dX\,dY}.
\end{equation}
When interaction term $S_{int}(\phi)$ is introduced in the Action quantum functional is conventionally given in a compact form by the expression $Z(j) = e^{S_{int}(\delta / \delta j)}\cdot Z_{free}(j)$. Thus in our case the replacement in bulk Witten diagrams of the Green functions $G$ with the Wightman difference $\widetilde G$ of two Green functions defined for regular and irregular b.c. means that instead of the standard quantum functional we consider:

\begin{equation}
\label{4}
{\widetilde Z} (j(X)) = e^{S_{int}(\delta / \delta j)}\cdot {\widetilde Z}_{free}(j),
\end{equation}
where ${\widetilde Z}_{free}(j)$ is given in (\ref{3}). Bulk diagrams built according to prescription (\ref{4}) may be called "Wightman-Witten diagrams". 

Using of the Wightman-Witten diagrams instead of the ordinary Witten diagram may be considered as just a toy trick that allows to carry out calculations to the visible results and to demonstrate in this way the possible effectiveness of the "old" conformal bootstrap approach in the AdS/CFT context. But perhaps this technique may be applied in other directions, in the "OPE conformal bootstrap" in particular? The excuse for such a drastic modification of the "first principles" may be in the fact that the whole AdS/CFT approach is not a consistent theory based on the immutable first principles, it remains so far a sort of Enigma with its main miracle of obtaining plethora of effects of quantum boundary theory in $d$ dimensions from the classical supergravity in $AdS_{d+1}$.

Actually Eq-s (\ref{1}) - (\ref{2}) are just a conventional Schwinger-Dyson equations written in planar approximation when "bare terms"are put equal to zero. This may be called the zero-Lagrangian approach applied by Sakharov in his quantum induced theory of gravity \cite{Sakharov}. The attempt to follow this way in the AdS context was made in \cite{Alt1}, \cite{Alt2} where UV-finite induced gravitational and gauge coupling constants were calculated. The field theory bootstrap is actually used in the mean-field theory of superconductivity, in dynamical symmetry breaking pioneered in \cite{Jona} and in many other aspects.  

The general motivation of all bootstrap efforts is a hope to calculate in this way the values of the fundamental constant, now introduced in the theory {\it{ad hoc}}. The goal of this paper is to demonstrate on the toy model that "hunting for numbers" of conformal dimensions may be successful when "old" conformal bootstrap is considered in the AdS/CFT context.

In Sec.2 familiar expressions used in the bulk of the paper are summed up. Sec. 3 describes self-energy bootstrap when primary conformal correlators are equated to the 2-point bubble Witten diagrams; two models are considered: (1) of three scalar fields  with untrivial spectra of their conformal dimensions as a result, and (2) $O(N)$-symmetric model of $N+1$ scalar fields which conformal dimensions satisfying bootstrap conditions acquire small imaginary parts. In Sec. 4 simple expressions are obtained for the four-point "Wightman-Witten" amplitudes that serve a Kernels in the Bethe-Salpeter type equations for the composite conformal operators considered in Sec. 5. The simple expressions for spectra of scaling dimension of these operators are obtained in Sec. 5. In Conclusion the possible directions of future work are outlined.

\section{Preliminaries}

\quad We work in $AdS_{d+1}$ in Poincare Euclidean coordinates $Z = \{z_{0}, {\vec z}\,\}$, where AdS curvature radius $R_{AdS}$ is put equal to one:

\begin{equation}
\label{5}
ds^{2} = \frac{dz_{0}^{2} + d {\vec z}\,^{2}}{z_{0}^{2}},
\end{equation}
and consider bulk scalar fields. Bulk field $\phi (X)$ of mass $m$ is dual to boundary conformal operator $O_{\Delta_{+}} (\vec x)$ or to its "shadow" operator $O_{\Delta_{-}}(\vec x)$ with scaling dimensions

\begin{equation}
\label{6}
\Delta_{\pm} = \frac{d}{2} \pm \sqrt{\frac{d^{2}}{4} + m^{2}} \equiv \frac{d}{2} \pm \nu.
\end{equation}

We take normalization of the scalar field bulk-to-boundary operator and of the corresponding conformal correlator like in \cite{Giombi1}, \cite{Giombi2}:

\begin{eqnarray}
\label{7}
K_{\Delta} (Z; \vec x) = \lim_{\stackrel {x_{0} \to 0}{}} \left[\frac{G_{\Delta}^{BB} (Z, X)}{(x_{0})^{\Delta}}\right] = C_{\Delta}\, \left [\frac{z_{0}}{z_{0}^{2} + (\vec z - \vec x)^{2}}\right]^{\Delta}, \nonumber
\\
\\
 C_{\Delta} = \frac{\Gamma (\Delta)}{2\pi^{d/2}\Gamma \left(1 + \Delta - \frac{d}{2}\right)}, \qquad  \qquad \qquad  \qquad \nonumber
\end{eqnarray}
and:

\begin{equation}
\label{8}
<O_{\Delta}({\vec x}) O_{\Delta} ({\vec y})> = \lim_{\stackrel{x_{0} \to 0}{y_{0} \to 0}} \left[\frac{G_{\Delta}^{BB} (X, Y)}{(x_{0}\,y_{0})^{\Delta}}\right]= \frac{C_{\Delta}}{P_{xy}^{\Delta}}, \, \, \,  P_{xy} = |{\vec x} - {\vec y}|^{2}
\end{equation}

Bulk-to-bulk scalar field Green function $G_{\Delta}^{BB} (X, Y)$ is given by the Kallen-Lehmann-split representation (see e.g. \cite{Giombi1}, \cite{Giombi2}, \cite{Penedones}, \cite{Fitz} and references therein):

\begin{eqnarray}
\label{9}
G_{\Delta}^{BB} (X, Y) = C_{\Delta}\,\left(\frac{\xi}{2}\right)^{\Delta}\,F\left( \frac{\Delta}{2}, \, \frac{\Delta + 1}{2}, \Delta - \frac{d}{2} + 1; \, \xi^{2} \right) = \nonumber
\\
\\
= \int_{-\infty}^{+\infty} \frac{c^{2}dc}{\pi[c^{2} + (\Delta - \frac{d}{2})^{2}]} \cdot \int \, K_{\frac{d}{2} + ic}(X; {\vec x}_{a})\, K_{\frac{d}{2} - ic}(Y; {\vec x}_{a})\, d^{d}{\vec x}_{a} \, , \nonumber
\end{eqnarray}

$$
\xi = \frac{2 \, x_{0} \, y_{0}}{x_{0}^{2} + y_{0}^{2} + ({\vec x} - {\vec y})^{2}}.
$$

However in this paper, following \cite{Giombi2} and \cite{Mitra} - \cite{Alt2}, the simplification of calculations is obtained with postulating the using of the "Wightman-Witten" quantum functional (\ref{4}) i.e. the changing in all Witten diagrams bulk-to-bulk Green functions (\ref{9}) to Wightman functions $\widetilde {G}_{\Delta} (X, Y)$ which is a difference of residues of poles $c = \pm i (\Delta - d/2)$ in the RHS of (\ref{9}):

\begin{equation}
\label{10}
{\widetilde G}_{\Delta}(X, Y) = G_{\Delta}^{BB} - G_{d-\Delta}^{BB} = (d - 2\Delta)\, \int \, K_{\Delta}(X; {\vec x}_{a})\, K_{d-\Delta}(Y; {\vec x}_{a})\, d^{d}{\vec x}_{a}.
\end{equation}
The choice of sign in this definition of $\widetilde G$ is essential. We took the sign like in \cite{Mitra} - \cite{Alt2} where tadpole ${\widetilde G}_{\Delta} (X, X)$ gave the correct sign of the UV-finite one-loop quantum potential. 

We shall also need expression for AdS/CFT tree vertex \cite{Freedman}, \cite{Paulos}, \cite{Giombi1}, \cite{Giombi2}; $g_{123}$ is a coupling constant of three bulk scalar fields:

\begin{eqnarray}
\label{11}
\Gamma_{\Delta_{1}, \Delta_{2}, \Delta_{3}} ({\vec x}_{1}, {\vec x}_{2}, {\vec x}_{3}) = 
g_{123}\int \,K_{\Delta_{1}} (X; {\vec x}_{1})\, K_{\Delta_{2}} (X; {\vec x}_{2}) \, K_{\Delta_{3}} (X; {\vec x}_{3}) \, dX =  \nonumber
\\  
\\
= g_{123}\, \frac{B(\Delta_{1}, \Delta_{2}, \Delta_{3})}{P_{12}^{\delta_{12}}\,P_{13}^{\delta_{13}}\, P_{23}^{\delta_{23}}}, \qquad  \, \qquad \, \qquad  \, \qquad \, \qquad  \nonumber
\end{eqnarray}
where

\begin{equation}
\label{12}
\delta_{12} = \frac{\Delta_{1} + \Delta_{2} - \Delta_{3}}{2}; \,\,\, \delta_{13} = \frac{\Delta_{1} + \Delta_{3} - \Delta_{2}}{2}; \,\,\, \delta_{23} = \frac{\Delta_{2} + \Delta_{3} - \Delta_{1}}{2},
\end{equation}

\begin{equation}
\label{13}
B(\Delta_{1}, \Delta_{2}, \Delta_{3}) = \frac{\pi^{d/2}}{2}\, \left( \prod\limits_{i=1}^{3}\frac{C_{\Delta_{i}}}{\Gamma (\Delta_{i})}\right) \cdot \Gamma \left(\frac{\Sigma \Delta_{i} - d}{2}\right)\cdot \Gamma(\delta_{12})\, \Gamma(\delta_{13}) \, \Gamma (\delta_{23}).
\end{equation}

Also some well known \cite{Symanzik2}, \cite{Parisi}, \cite{Fradkin}, \cite{Giombi1}, \cite{Giombi2} conformal integrals will be used below:

\begin{equation}
\label{14}
\int \frac{d^{d}{\vec y}}{P_{1y}^{\Delta_{1}} \,P_{2y}^{\Delta_{2}}\, P_{3y}^{\Delta_{3}}} \stackrel {\Sigma \Delta_{i} = d} {=} \frac{A(\Delta_{1}, \Delta_{2}, \Delta_{3})}{P_{12}^{\frac{d}{2} - \Delta_{3}}\,P_{13}^{\frac{d}{2} - \Delta_{2}}\, P_{23}^{\frac{d}{2} - \Delta_{1}}},
\end{equation}
and

\begin{equation}
\label{15}
\int \frac{d^{d}{\vec y}}{P_{1y}^{\Delta_{1}} \,P_{2y}^{\Delta_{2}}} = \frac{A(\Delta_{1}, \Delta_{2}, d - \Delta_{1} - \Delta_{2})} {P_{12}^{\Delta_{1} + \Delta_{2} - \frac{d}{2}}},
\end{equation}
where

\begin{equation}
\label{16}
A(\Delta_{1}, \Delta_{2}, \Delta_{3}) = \frac{\pi^{d/2}\, \Gamma (\frac{d}{2} - \Delta_{1}) \, \Gamma (\frac{d}{2} - \Delta_{2}) \, \Gamma (\frac{d}{2} - \Delta_{3})}{\Gamma (\Delta_{1}) \, \Gamma(\Delta_{2}) \, \Gamma (\Delta_{3})}.
\end{equation}

We will also need integral (\ref{15}) when $\Delta_{1} + \Delta_{2} = d$, $\Delta_{1} \ne \Delta_{2} \ne d/2$ (the derivation of this formula is elementary in momentum space, see e.g. in \cite{Fradkin}):

\begin{equation}
\label{17}
\int \frac{d^{d}{\vec y}}{P_{1y}^{\Delta} \,P_{2y}^{d - \Delta}} = \frac{\pi^{d}\,\Gamma (\Delta - \frac{d}{2})\, \Gamma (\frac{d}{2} - \Delta)}{\Gamma (\Delta) \, \Gamma (d - \Delta)} \cdot \delta^{(d)} ({\vec x}_{1} - {\vec x}_{2}), \qquad \Delta \ne \frac{d}{2};
\end{equation}

and the divergent integral analyzed in detail in \cite{Giombi1}:

\begin{equation}
\label{18}
\int \frac{d^{d}{\vec y}}{P_{1y}^{\frac{d}{2}} \,P_{2y}^{\frac{d}{2}}} = \frac{A(\frac{d}{2}, \frac{d}{2}, 0)}{P_{12}^{\frac{d}{2}}} = \frac{\pi^{d/2} \, \Gamma (0)}{\Gamma(\frac{d}{2})} \cdot \frac{1}{P_{12}^{\frac{d}{2}}},
\end{equation}
(we used here (\ref{15}) and (\ref{16}) for $\Delta_{1} = \Delta_{2} = d/2$).

In (\ref{18}) $\Gamma (0)$ is written symbolically, its dimensional regularization is presented in \cite{Giombi1}. The point is that, like it was observed in \cite{Giombi2}, this divergence reduces in final expressions for Wightman-Witten diagrams if the "extremal" vertexes are considered and the norm-invariant squared coupling constant is used. Definition of these notions see below.

Throughout the paper two models are considered:

I. Model of three bulk scalar fields $\phi (X)$, $\psi (X)$ and $\sigma (X)$ with their bulk interaction 

\begin{equation}
\label{19}
L^{I}_{int} = g_{I} \, \phi (X) \psi (X) \sigma (X).
\end{equation}
These bulk fields correspond to the boundary conformal operators $O_{\Delta_{\phi}}$, $O_{\Delta_{\psi}}$, $O_{\Delta_{\sigma}}$ with corresponding conformal dimensions $\Delta_{\phi}$, $\Delta_{\psi}$ and $\Delta_{\sigma}$ of the conformal correlators (\ref{8}), or to their "shadow" counterparts with conformal dimensions $d - \Delta_{\phi}$, $d - \Delta_{\psi}$ and $d - \Delta_{\sigma}$.

II. $O(N)$ symmetric model of $N + 1$ scalar fields $\psi_{i}(X)$ $(i = 1...N)$ and singlet field $\sigma (X)$ that is conventionally obtained with the Hubbard-Stratonovich transformation from the model of $N$ fields $\psi_{i}$ with the $O(N)$ symmetric quartic interaction $[\Sigma_{i} \psi_{i}^{2} (X)]^{2}$. Conformal dimensions of fields $\psi_{i}$ are equal:

$$
\Delta_{\psi_{1}} = ... \Delta_{\psi_{N}} \equiv {\widetilde \Delta}_{\psi},
$$
and the bulk interaction of the model is

\begin{equation}
\label{20}
L^{II}_{int} = g_{II} \, \Sigma_{i} \psi_{i}^{2} (X) \, \sigma (X).
\end{equation}

$g_{I,\,II}$ in (\ref{19}), (\ref{20}) are the "bare" coupling constants, that, according to \cite{Giombi2}, must be renormalized. Namely in Witten diagrams $g_{I,\,II}$ squared must be changed to the norm-invariant:

\begin{eqnarray}
\label{21}
g^{* 2}_{I} = g^{2}_{I} \cdot a^{\rm{UV}}_{I} = g^{2}_{I} \cdot \frac{B^{2}(\Delta_{\phi}, \Delta_{\psi}, \Delta_{\sigma})}{C_{\Delta_{\phi}} C_{\Delta_{\psi}} C_{\Delta_{\sigma}}},  \qquad \nonumber
\\
\\
g^{* 2}_{II} = g^{2}_{II} \cdot a^{\rm{UV}}_{II} = g^{2}_{II} \cdot \frac{B^{2}({\widetilde \Delta}_{\psi}, d - {\widetilde \Delta}_{\psi}, \Delta_{\sigma})}{C_{{\widetilde \Delta}_{\psi}} C_{d - {\widetilde \Delta}_{\psi}} C_{\Delta_{\sigma}}},   \nonumber
\end{eqnarray}
here $B (\Delta_{1}, \Delta_{2}, \Delta_{3})$ and $C_{\Delta}$ are given in (\ref{13}) and (\ref{7}).

In \cite{Giombi2} it is observed that Witten diagrams become easily calculable in case three-point vertexes are "extremal" \cite{extremal}. In our case this means that conformal dimensions corresponding to fields $\phi$, $\psi$, $\sigma$ of the model I obey:

\begin{equation}
\label{22}
\Delta_{\phi} = \Delta_{\psi} + \Delta_{\sigma},
\end{equation}
whereas in model II field $\phi$ of model I in (\ref{22}) must be changed to the shadow counterparts of the fields $\psi_{i}$ of conformal dimension ${\widetilde \Delta}_{\psi}$, then extremal condition (\ref{22}) looks as:

\begin{equation}
\label{23}
d - {\widetilde \Delta}_{\psi} = {\widetilde \Delta}_{\psi} + \Delta_{\sigma},
\end{equation}

In case (\ref{22}), (\ref{23}) are valid norm-invariant coefficients $a^{\rm{UV}}_{I,\,II}$ (\ref{21}) include $\Gamma ^{2} (0)$ in numerator and according to \cite{Giombi2} this infinity is absorbed in "bare" $g^{2}_{I,\,II}$. Then, as it is shown in \cite{Giombi2}, Witten diagrams expressed through renormalized squared coupling constant $g^{* 2}_{I,\,II}$ (\ref{21}) are finite and are extremely simple.

\section{AdS bubble analogy of the Green function bootstrap}

\vspace{0,5cm}

\qquad {\bf{\Large{\it Model I}}}.

\vspace{0,5cm}

\qquad Let us consider the following AdS/CFT version of the first of the field theory symbolic Eq-s (\ref{1}) written for the boundary conformal correlator (\ref{8}) of field $\phi$ whereas bubble ("self-energy" of $\phi$) is built of fields $\psi$, $\sigma$:

\begin{equation}
\label{24}
<O_{\phi}({\vec x}_{1}) O_{\phi} ({\vec x}_{2})> \, = \, {\widetilde M}^{{\rm 2pt \,\, bubble}}_{\Delta_{\phi}},
\end{equation}
where LHS is given in (\ref{8}), and in calculation of Witten bubble diagram in the RHS we follow \cite{Giombi2} with substitution, according to procedure (\ref{4}), Wightman functions (\ref{10}) instead of bulk-to-bulk Green functions (\ref{9}).

In model I with the bulk interaction (\ref{19}) bootstrap Eq. (\ref{24}) looks as:

\begin{equation}
\label{25}
\frac{C_{\Delta_{\phi}}}{P_{12}^{\Delta_{\phi}}} = g^{2}_{I} \int K_{\Delta_{\phi}}(X ; {\vec x}_{1})\,{\widetilde G}_{\Delta_{\psi}}(X, Y)\,{\widetilde G}_{\Delta_{\sigma}} (X, Y)\,K_{\Delta_{\phi}}(Y ; {\vec x}_{2})\, dX\,dY.
\end{equation}

Actually this is nothing but first of symbolic equations (\ref{1}) written for bulk Green function (\ref{9}) where "external" points $x_{1}$, $x_{2}$ are put on the AdS boundary; then $G$ in the LHS of (\ref{1}) becomes correlator (\ref{8}) in the LHS of (\ref{25}), whereas two $G$ on the RHS of (\ref{1}) become bulk-to-boundary operators (\ref{7}) in the RHS of (\ref{25}) \footnote{I am grateful to Ruslan Metsaev for this observation.}.

Taking ${\widetilde G} (X, Y)$ from (\ref{10}) and performing two bulk integrals (see (\ref{11})-(\ref{13})) gives for integral in the RHS of (\ref{25}):

\begin{eqnarray}
\label{26}
(d-2\Delta_{\psi})(d- 2\Delta_{\sigma}) \int d{\vec x}_{a}d{\vec x}_{b} \int \, K_{\Delta_{\phi}}(X ; {\vec x}_{1})\,K_{\Delta_{\psi}}(X ; {\vec x}_{a})\,K_{\Delta_{\sigma}}(X ; {\vec x}_{b})\,dX\, \cdot  \qquad  \nonumber 
\\   \nonumber
\\  \nonumber
\\ 
\cdot \int\, K_{d - \Delta_{\psi}}(Y ; {\vec x}_{a})\,K_{d- \Delta_{\sigma}}(Y ; {\vec x}_{b})\,K_{\Delta_{\phi}}(Y ; {\vec x}_{2})\,dY =    \qquad \qquad \qquad  \qquad
\\ \nonumber
\\ \nonumber
= \int d{\vec x}_{a}d{\vec x}_{b} \,\left[\frac{(d-2\Delta_{\psi})(d- 2\Delta_{\sigma})\, B(\Delta_{\phi}, \Delta_{\psi}, \Delta_{\sigma})\, B(\Delta_{\phi}, d- \Delta_{\psi}, d - \Delta_{\sigma})}{P_{1a}^{\delta_{1a}} \, P_{1b}^{\delta_{1b}}\,P_{ab}^{\delta^{(1)}_{ab}}\, P_{2a}^{\delta_{2a}} \, P_{2b}^{\delta_{2b}}\,P_{ab}^{\delta^{(2)}_{ab}} }\right], \qquad  \nonumber
\end{eqnarray}
where

\begin{eqnarray}
\label{27}
\delta_{1a} = \delta_{2b} = \frac{\Delta_{\phi} + \Delta_{\sigma} - \Delta_{\psi}}{2}; \,\,\, \delta_{1b} = \delta_{2a} = \frac{\Delta_{\phi} + \Delta_{\psi} - \Delta_{\sigma}}{2}; \nonumber
\\
\\
\delta^{(1)}_{ab} = \frac{\Delta_{\psi} + \Delta_{\sigma} - \Delta_{\phi}}{2}; \, \,\,\,\,\,\, \delta^{(2)}_{ab} = \frac{2d - \Delta_{\phi} - \Delta_{\psi} - \Delta_{\sigma}}{2}.  \quad \nonumber
\end{eqnarray}

Conformal integrals in (\ref{26}) may be taken with the help of standard formulas (\ref{14}) - (\ref{18}). Following \cite{Giombi2} we use in (\ref{26}) instead of $g^{2}_{I}$ norm-invariant squared coupling constant $g^{* 2}_{I} = g^{2}_{I}\, a^{\rm{UV}}_{I}$ (\ref{21}). Then (\ref{25}) takes the form:

\begin{eqnarray}
\label{28}
\frac{C_{\Delta_{\phi}}}{P_{12}^{\Delta_{\phi}}} = g^{* 2}_{\phi \psi \sigma} \, C_{\Delta_{\phi}}\,C_{\Delta_{\psi}}\,C_{\Delta_{\sigma}} \,\frac{B(\Delta_{\phi}, d - \Delta_{\psi}, d - \Delta_{\sigma})}{B(\Delta_{\phi}, \Delta_{\psi}, \Delta_{\sigma})} \, \cdot \nonumber
\\
\\
\cdot \, (d - 2\Delta_{\psi})\, (d - 2\Delta_{\sigma})\,\frac{A(\delta_{1b}, \delta_{2b}, d - \Delta_{\phi}) \, A (\frac{d}{2}, \frac{d}{2}, 0)}{P_{12}^{\Delta_{\phi}}},    \nonumber
\end{eqnarray}
expressions for $C$, $B$ and $A$ see in (\ref{7}), (\ref{13}), (\ref{16}), (\ref{18}).

In case of validity of the "extremal" equality (\ref{22}) between the conformal dimensions $\Delta_{\phi}$, $\Delta_{\psi}$, $\Delta_{\sigma}$ the infinity - $\Gamma (0)$ in $A (d/2, d/2, 0)$ in the RHS of (\ref{28}) is reduced with the similar infinity in $B(\Delta_{\phi}, \Delta_{\psi}, \Delta_{\sigma})$. This important observation in \cite{Giombi2} permitted to obtain the finite sensible answers.

Also the validity of (\ref{22}) leads to the reduction of practically all $\Gamma$-functions in the RHS of (\ref{28}). The remarkable simplicity of expressions for conformal correlators in case of the "extremal" relation between conformal dimensions was observed in \cite{Giombi2}.

Here are the final forms of the bubble bootstrap equation (\ref{25}) (or (\ref{28})) for the correlation function of field $\phi$, and of similar to (\ref{25}) equations for the correlation functions corresponding to fields $\psi$, $\sigma$ when bubbles are formed of fields $(\phi, \sigma)$ and $(\phi, \psi)$ correspondingly:

\begin{equation}
\label{29}
\frac{C_{\Delta_{\phi}}}{P_{12}^{\Delta_{\phi}}} = \frac{g^{* 2}_{I} \, C_{\Delta_{\phi}}}{P_{12}^{\Delta_{\phi}}},  \qquad  \qquad \,  \,
\end{equation}

\begin{equation}
\label{30}
\frac{C_{\Delta_{\psi}}}{P_{12}^{\Delta_{\psi}}} = \frac{g^{* 2}_{I} \, C_{\Delta_{\psi}}}{P_{12}^{\Delta_{\psi}}} \, \cdot \, \frac{F(\Delta_{\phi})}{F (\Delta_{\psi})},
\end{equation}

\begin{equation}
\label{31}
\frac{C_{\Delta_{\sigma}}}{P_{12}^{\Delta_{\sigma}}} = \frac{g^{* 2}_{I} \, C_{\Delta_{\sigma}}}{P_{12}^{\Delta_{\sigma}}} \, \cdot \, \frac{F(\Delta_{\phi})}{F (\Delta_{\sigma})},
\end{equation}
where

\begin{equation}
\label{32}
F (\Delta_{i}) = \frac{\Gamma (\Delta_{i})\, \Gamma(d - \Delta_{i})}{\Gamma(\Delta_{i} - \frac{d}{2}) \, \Gamma(\frac{d}{2} - \Delta_{i})}  \, \, \, \, \, \, (i = \phi, \, \psi, \, \sigma).
\end{equation}

Thus Eq. (\ref{29}) fixes the norm-invariant coupling constant $g^{*2}_{I}$ (\ref{21}) (in units $R_{AdS}^{(d - 5)}$):

\begin{equation}
\label{33}
g^{* 2}_{I} = 1,
\end{equation}
and (\ref{30})-(\ref{32}), (\ref{22}) give a system determining spectra of conformal dimensions:

\begin{equation}
\label{34}
F(\Delta_{\phi}) = F(\Delta_{\psi}) = F(\Delta_{\sigma}); \,\,\,\,  \Delta_{\phi} = \Delta_{\psi} + \Delta_{\sigma}.
\end{equation}

Eq-s (\ref{33}), (\ref{34}) are the main result of this Section for model I. We pay attention that according to (\ref{32}) $F(\Delta) = F(d - \Delta)$, hence self-energy bootstrap of shadow fields again gives spectral Eq-s (\ref{34}).

Function $F(\Delta)$ (\ref{32}) for even $d$ is a polynomial of degree $d$, whereas for odd $d$ it is expressed in terms of trigonometric functions. Here are functions $F(\Delta)$ for $d = 1, 2, 3, 4$:

\begin{eqnarray}
\label{35}
F_{d=1}(\Delta) = \frac{(\Delta - 1/2) \, \cos\pi\Delta}{\sin\pi\Delta}; \qquad F_{d=4} = (\Delta - 1)(\Delta - 2)^{2}(\Delta - 3);   \nonumber
\\
\\
F_{d=2}(\Delta) = - (\Delta - 1)^{2}; \,\,F_{d=3}(\Delta) =  - \frac{(\Delta - 1) (\Delta - 3/2) (\Delta - 2)\, \cos\pi\Delta}{\sin\pi\Delta}.  \nonumber
\end{eqnarray}

Spectral Eq-s (\ref{34}) possess finite number of roots for even $d$ and infinite number of roots for odd $d$. However if the condition of unitary bound $(d - 2) < 2\,\Delta_{-} < d$, $\Delta_{-}$ see in (\ref{6}), is imposed the number of solutions of (\ref{34}) is limited. Let us look at some of them.

The evident one is valid for any dimension $d$:

\begin{equation}
\label{36}
\Delta_{\psi} = \Delta_{\sigma} = \frac{d}{3} \qquad \,\,\, \Delta_{\phi} = d - \Delta_{\psi} = \frac{2\,d}{3}.
\end{equation}
It means that in this case three bulk fields $\phi$, $\psi$, $\sigma$ are actually one and the same self-interacting bulk field of $m^{2} = \Delta (\Delta - d) = - 2 d^{2} / 9$ (in units $R_{AdS}^{-2}$), that is $\Delta_{\phi} = \Delta_{+}$ and $\Delta_{\psi} = \Delta_{\sigma} = \Delta_{-}$ for this $m^{2}$. $\nu = d/6$ in this case, see (\ref{6}), thus unitary bound for solution (\ref{36}) is valid for $d < 6$.

For $d = 4$ there are three more solutions obtained from (\ref{34}), (\ref{35}):

The first one:

\begin{equation}
\label{37}
\Delta_{\phi} = \frac{14}{5}, \qquad \,\,\, \Delta_{\psi} = \Delta_{\sigma} = \frac{7}{5},
\end{equation}
that corresponds to $m^{2}_{\phi} = - 84/25$ ($\Delta_{\phi} = \Delta_{+}$ for this $m^{2}$ and $d = 4$ in (\ref{6})) and to $m^{2}_{\psi} = m^{2}_{\sigma} = - 91/25$ ($\Delta_{\psi} = \Delta_{\sigma} = \Delta_{-}$ for this $m^{2}$ in (\ref{6})).

In the second solution field $\phi$ is a "shadow" of $\psi$:

\begin{equation}
\label{38}
\Delta_{\phi} = \frac{13}{5}, \qquad \,\, \Delta_{\psi} = \frac{7}{5}, \qquad \,\, \Delta_{\sigma} = \frac{6}{5}.
\end{equation}
This corresponds to $m^{2}_{\phi} = m^{2}_{\psi} = -91/25$ ($\Delta_{\phi} = \Delta_{+}$, $\Delta_{\psi} = d - \Delta_{\phi} = \Delta_{-}$ in (\ref{6})), and $m^{2}_{\sigma} = -84/25$ ($\Delta_{\sigma} = \Delta_{-}$ in (\ref{6})).

The last solution of the spectral Eq-s (\ref{34}), (\ref{35}) for $d = 4$ is obtained with the permutation of $\psi$ and $\sigma$ in (\ref{38}).

If $\phi$ is a shadow of $\psi$ (solutions (\ref{36}), (\ref{38})) spectral Eq-s (\ref{34}), with account of (\ref{23}), come to the simple condition:

\begin{equation}
\label{39}
F(\Delta_{\psi}) = F(2\,\Delta_{\psi}),
\end{equation}
where $F(\Delta)$ is given in (\ref{32}) or (\ref{35}). Surely Eq. (\ref{39}) has solutions (\ref{36}) and (\ref{38}), in particular $\Delta_{\psi} = 2/3$ for $d=2$ and $\Delta_{\psi} = 4/3$ or $7/5$ for $d = 4$. For $d = 1$ and 
$d = 3$ (\ref{39}) and (\ref{35}) give following spectral equations:

\begin{eqnarray}
\label{40}
d = 1: \qquad \tan^{2}\pi\Delta_{\psi} = \frac{1}{4\,\Delta_{\psi} - 1}; \qquad  \quad  \nonumber
\\
\\ 
d = 3: \qquad \tan^{2}\pi\Delta_{\psi} = \frac{3\,(\Delta_{\psi} - 1)\,(2\,\Delta_{\psi} + 1)}{(4\,\Delta_{\psi} - 3)\,(2\,\Delta_{\psi} -1)}.   \nonumber
\end{eqnarray}

For $d = 1$ there is only one solution (\ref{36}) (that is $\Delta_{\psi} = \Delta_{\sigma} = 1/3$, $\Delta_{\phi} = 2/3$) satisfying unitary bound condition $-1/2 < \Delta_{-} < 1/2$. 

Whereas for $d = 3$ unitary bound $1/2 < \Delta_{-} < 3/2$ is valid for two solutions of (\ref{39}), (\ref{40}): 

\begin{equation}
\label{41}
(1) \Delta_{\psi} = \Delta_{\sigma} = 1, \,\, \Delta_{\phi} = 2; (2) \Delta_{\psi} = 1,24, \,\, \Delta_{\sigma} = 0,55, \,\, \Delta_{\phi} = 1,76.
\end{equation}

\vspace{0,5cm}

{\bf{\Large{\it Model II}}}.

\vspace{0,5cm}

This $O(N)$ symmetric model II of $N$ identical scalar fields $\psi_{i}$, interacting  with singlet scalar field $\sigma$ according to (\ref{20}), in case $N = 1$ evidently coincides with model I if field $\phi$ is a shadow of $\psi$ when spectral rule (\ref{39}) and solutions (\ref{36}), (\ref{38}), (\ref{41}) are valid.

For $N > 1$ bootstrap self-energy condition for fields $\psi_{i}$ is not changed as compared to Eq-s (\ref{25}), (\ref{28}) - (\ref{30}), (\ref{32}) of model I (with account of $F({\widetilde\Delta}_{\psi}) = F(d - {\widetilde\Delta}_{\psi})$) because bubble of the field $\psi_{i}$ is formed with the same field $\psi_{i}$ and field $\sigma$. Thus first bootstrap condition of model I $g^{* 2}_{I} = 1$ (\ref{33}) is preserved in model II: $g^{* 2}_{II} = 1$ (definition of $g^{*}_{II}$ see in (\ref{20}), (\ref{21})).

However bootstrap self-energy condition for field $\sigma$ in model II:

\begin{equation}
\label{42}
\frac{C_{\Delta_{\sigma}}}{P_{12}^{\Delta_{\sigma}}} = \frac{g^{* 2}_{II} \, C_{\Delta_{\sigma}}}{P_{12}^{\Delta_{\sigma}}} \, \cdot \, N\, \frac{F({\widetilde \Delta}_{\psi})}{F (\Delta_{\sigma})}
\end{equation}
now acquires factor $N$ as compared to (\ref{31}) since according to (\ref{20}) $\sigma$ interacts with every of $N$ fields $\psi_{i}$, and its bubble is formed with all $\psi_{i}$. 

Finally taking into account that $g^{* 2}_{II} = 1$, relations $\Delta_{\sigma} = d - 2\,{\widetilde \Delta}_{\psi}$ (cf. \ref{23}) and $F(\Delta_{\sigma}) = F(d - \Delta_{\sigma})$ (see (\ref{32})) the following bootstrap spectral condition in model II is obtained from (\ref{42}) instead of Eq. (\ref{39}) of model I:

\begin{equation}
\label{43}
N \, F({\widetilde \Delta}_{\psi}) = F(2\, {\widetilde \Delta}_{\psi}).
\end{equation}

Explicit expressions (\ref{35}) of function $F(\Delta)$ for different dimensions $d$ permit to get solutions of (\ref{43}). For $d = 4$ there is always the unphysical solution valid for any $N$: ${\widetilde \Delta}_{\psi} = 1$, when $F_{d=4}(1) = F_{d=4}(2) = 0$. Other roots of (\ref{43}) for $d = 4$ and $N = 2, 3, 4$ are:

\begin{eqnarray}
\label{44}
N = 2: \,\,\, {\widetilde \Delta}_{\psi} = 1.4235 \pm i\,0.1486; \,\,\, -\,0.4184   \nonumber
\\
N=3: \,\,\, {\widetilde \Delta}_{\psi} = 1.4628 \pm i\,0.1883; \,\,\, -\,0.8487
\\
N = 4: \,\,\, {\widetilde \Delta}_{\psi} = 1.4931 \pm i\,0.2104; \,\,\, -\,1.3195   \nonumber
\end{eqnarray} 
Then conformal dimension of field $\sigma$ is received from the "extremal" relation (\ref{23}) $\Delta_{\sigma} = 4 - 2\, {\widetilde \Delta}_{\psi}$.

Discussion of possible physical applications of this result is beyond the scope of this article. We note only that complex conformal dimensions are not something unusual. As a rule it signals about the nonunitarity of the theory, see e.g. \cite{Hogervorst}. However in the recent paper \cite{Metsaev} it is argued that complexity of conformal dimensions guarantees classical unitarity of the continuous-spin field.

\section{Four-point "extremal" correlators}

\qquad In the next section the AdS/CFT analogy of the Bethe-Salpeter field theory bootstrap equations (second symbolic eq. in (\ref{1})) will be considered in the model I of three scalar fields with interaction (\ref{19}). In this section the Kernels of these equations are calculated. Here the Kernels are the connected one-channel tree "Wightman-Witten" diagrams for the four-point correlators ${\vec x}_{1}{\vec x}_{2} \to {\vec x}_{a}{\vec x}_{b}$ of fields $\psi$, $\phi$, $\sigma$ or of their "shadows". For example:

\begin{equation}
\label{45}
{\widetilde M}^{\rm{4pt\, tree}}_{\Delta_{\psi},\,\Delta_{\phi}\,|\,\Delta_{\phi},\, \Delta_{\psi}} ({\vec x}_{1}, {\vec x}_{2}, {\vec x}_{a}, {\vec x}_{b})= \, <O_{\Delta{\psi}}({\vec x}_{1})\, O_{\Delta{\phi}}({\vec x}_{2}) \, O_{\Delta_{\phi}}({\vec x}_{a}) \, O_{\Delta_{\psi}}({\vec x}_{b})>, 
\end{equation}
where interaction is carried out by means of the field $\sigma (X)$ in the "$t$-channel" ${\vec x}_{1}{\vec x}_{a} \to {\vec x}_{2}{\vec x}_{b}$, and bulk Green function $G^{BB}_{\Delta_{\sigma}} (X, Y)$ (\ref{9}) is changed to corresponding Wightman function ${\widetilde G}^{BB}_{\Delta_{\sigma}} (X, Y)$ (\ref{10}):

\begin{eqnarray}
\label{46}
{\widetilde M}^{\rm{4pt\, tree}}_{\Delta_{\psi}, \,\Delta_{\phi}\,|\,\Delta_{\phi},\, \Delta_{\psi}} ({\vec x}_{1}, {\vec x}_{2}, {\vec x}_{a}, {\vec x}_{b})= g^{2}_{I} \cdot \qquad  \qquad   \qquad  \qquad \nonumber
\\
\\
\cdot \int\,dX\,dY\, K_{\Delta_{\psi}}(X;{\vec x}_{1})\, K_{\Delta_{\phi}}(X;{\vec x}_{a}) \, {\widetilde G}_{\Delta_{\sigma}} (X, Y)\, K_{\Delta_{\phi}}(Y;{\vec x}_{2})\, K_{\Delta_{\psi}}(Y;{\vec x}_{b}).  \nonumber
\end{eqnarray}

The subsequent computation algorithm is obvious:

- to substitute in (\ref{46}) ${\widetilde G}_{\Delta_{\sigma}}(X,Y)$ from (\ref{10});

- to perform two bulk integrals using (\ref{11})-(\ref{13});

- to substitute "extremal" equality (\ref{22}) and then to perform boundary conformal integral using (\ref{14}) and (\ref{16});

- instead of the "bare" squared coupling constant $g^{2}_{I}$ to insert the norm-invariant one (\ref{21}).

Omitting these elementary steps we present the simple final expressions for three amplitudes ${\widetilde M}^{\rm{4pt\,tree}}$, symbolically: $\psi \phi \to \phi \psi$ (intermediate field in $t$-channel is $\sigma$), $\sigma \phi \to \phi \sigma$ (intermediate field $\psi$), $\psi \sigma \to \sigma_{\rm{shad}} \psi_{\rm{shad}}$ (intermediate field $\phi$). The first one is deciphered in (\ref{45}), two others are correlators 

$$
{\widetilde M}^{\rm{4pt\, tree}}_{\Delta_{\sigma},\,\Delta_{\phi}\,|\,\Delta_{\phi},\, \Delta_{\sigma}} ({\vec x}_{1}, {\vec x}_{2}, {\vec x}_{a}, {\vec x}_{b})= \, <O_{\Delta{\sigma}}({\vec x}_{1})\, O_{\Delta{\phi}}({\vec x}_{2}) \, O_{\Delta_{\phi}}({\vec x}_{a}) \, O_{\Delta_{\sigma}}({\vec x}_{b})>
$$

and

$$
{\widetilde M}^{\rm{4pt\, tree}}_{\Delta_{\psi},\,\Delta_{\sigma}\,|\,d - \Delta_{\sigma}, \, d - \Delta_{\psi}} ({\vec x}_{1}, {\vec x}_{2}, {\vec x}_{a}, {\vec x}_{b})= \, <O_{\Delta_{\psi}}({\vec x}_{1})\, O_{\Delta_{\sigma}}({\vec x}_{2}) \, O_{d - \Delta_{\sigma}}({\vec x}_{a}) \, O_{d - \Delta_{\psi}}({\vec x}_{b})>.
$$

Finally:

\begin{equation}
\label{47}
{\widetilde M}^{\rm{4pt\, tree}}_{\Delta_{\psi},\,\Delta_{\phi}\,|\,\Delta_{\phi},\, \Delta_{\psi}} ({\vec x}_{1}, {\vec x}_{2}, {\vec x}_{a}, {\vec x}_{b})= 
g^{*2}_{I} \cdot \frac{C_{\Delta_{\phi}}C_{\Delta_{\psi}}}{P_{1a}^{\Delta_{\psi}}\, P_{2b}^{\Delta_{\psi}} \, P_{2a}^{\Delta_{\sigma}}}, 
\end{equation}

\begin{equation}
\label{48}
{\widetilde M}^{\rm{4pt\, tree}}_{\Delta_{\sigma},\,\Delta_{\phi}\,|\,\Delta_{\phi},\, \Delta_{\sigma}} ({\vec x}_{1}, {\vec x}_{2}, {\vec x}_{a}, {\vec x}_{b})= , 
g^{*2}_{I} \cdot \frac{C_{\Delta_{\phi}}C_{\Delta_{\sigma}}}{P_{1a}^{\Delta_{\sigma}}\, P_{2b}^{\Delta_{\sigma}} \, P_{2a}^{\Delta_{\psi}}},
\end{equation}

\begin{eqnarray}
\label{49}
{\widetilde M}^{\rm{4pt\, tree}}_{\Delta_{\psi},\,\Delta_{\sigma}\,|\,d - \Delta_{\sigma}, \, d - \Delta_{\psi}} ({\vec x}_{1}, {\vec x}_{2}, {\vec x}_{a}, {\vec x}_{b})=
g^{*2}_{I} \cdot \frac{C_{\Delta_{\psi}}C_{\Delta_{\sigma}}}{P_{1a}^{\Delta_{\psi}}\, P_{2b}^{\Delta_{\sigma}} \, P_{ab}^{d - \Delta_{\phi}}} \, \cdot \nonumber
\\
\cdot \, F (\Delta_{\phi}) \, \cdot \, \frac{\Gamma(\Delta_{\psi} - \frac{d}{2}) \, \Gamma(\Delta_{\sigma} - \frac{d}{2}) }{\Gamma (\Delta_{\psi}) \, \Gamma(\Delta_{\sigma})}, \qquad  \qquad \qquad \qquad
\end{eqnarray}
$F(\Delta_{\phi})$ is defined in (\ref{32}).

\section{Spectra of "Wightman-Witten" four-point correlators and vertex "old" bootstrap}

\qquad Let us look at the 3-point correlator (\ref{11})

\begin{equation}
\label{50}
\Gamma_{\Delta_{\psi}, \Delta_{\phi}, \Delta_{(\psi\phi)}} ({\vec x}_{1}, {\vec x}_{2}, {\vec x}_{3}) = \, <O_{\psi}({\vec x}_{1})\, O_{\phi}({\vec x}_{2}) \, O_{\Delta_{(\psi\phi)}}({\vec x}_{(3)})>,
\end{equation}
where $O_{\psi}$, $O_{\phi}$ are primary conformal operators considered in previous sections and $O_{\Delta_{(\psi\phi)}}$ is a composite conformal operator. Bootstrap equation proposed below give the spectrum of conformal dimension $\Delta_{(\psi\phi)}$.

In the analogy with second symbolic field theory bootstrap equation in (\ref{1}) the equality of 3-point tree vertex (\ref{50}) to the corresponding three-point one-loop triangle Wightman-Witten diagram is proposed as a candidate for AdS/CFT vertex bootstrap equation:

\begin{eqnarray}
\label{51}
\Gamma_{\Delta_{\psi}, \Delta_{\phi}, \Delta_{(\psi\phi)}} ({\vec x}_{1}, {\vec x}_{2}, {\vec x}_{3}) = {\widetilde M}^{\rm{3pt \, triangle}}_{\Delta_{\phi},\Delta_{\psi}, \Delta_{(\psi\phi)}} ({\vec x}_{1}, {\vec x}_{2}, {\vec x}_{3}) = g^{2}_{I}\, g_{\phi \psi (\psi\phi)}\,\, \int dXdYdZ \cdot \nonumber
\\
\\
\cdot \{ K_{\Delta_{\psi}}(X;{\vec x}_{1})\,{\widetilde G}_{\Delta_{\sigma}}(X,Y)\,K_{\Delta_{\phi}}(Y;{\vec x}_{2}) \, {\widetilde G}_{\Delta_{\psi}}(Y,Z)\,K_{\Delta_{(\psi\phi)}}(Z;{\vec x}_{3})\,{\widetilde G}_{\Delta_{\phi}}(Z,X) \}. \nonumber
\end{eqnarray}
Here $g_{\phi \psi (\psi\phi)}$ is a bulk coupling constant of two primary fields $\phi (X)$, $\psi (X)$ and bulk scalar field corresponding to composite conformal operator $O_{\Delta_{(\psi\phi)}}$; it will reduce in final expressions since $\Gamma_{\Delta_{\psi}, \Delta_{\phi}, \Delta_{(\psi\phi)}} ({\vec x}_{1}, {\vec x}_{2}, {\vec x}_{3})$ in the LHS of (\ref{51}) includes it (see (\ref{11})).

Taking again ${\widetilde G}$ from (\ref{10}), using (\ref{11})-(\ref{13}) to perform three integrals over bulk coordinates $X, Y, Z$, and (\ref{14}), (\ref{16}), (\ref{22}) to perform integral over one of three boundary points we obtain from (\ref{51}) following Bethe-Salpeter equation:

\begin{eqnarray}
\label{52}
\Gamma_{\Delta_{\psi}, \Delta_{\phi}, \Delta_{(\psi\phi)}} ({\vec x}_{1}, {\vec x}_{2}, {\vec x}_{3}) =  \int \, {\widetilde M}^{\rm{4pt\, tree}}_{\Delta_{\psi},\,\Delta_{\phi}\,|\,\Delta_{\phi},\, \Delta_{\psi}} ({\vec x}_{1}, {\vec x}_{2}, {\vec x}_{a}, {\vec x}_{b}) \cdot \nonumber
\\
\\
\cdot \, (d - 2\Delta_{\phi}) \, (d - 2\Delta_{\psi}) \, \Gamma_{d - \Delta_{\phi}, d - \Delta_{\psi}, \Delta_{(\psi\phi)}} ({\vec x}_{a}, {\vec x}_{b}, {\vec x}_{3}) \, d{\vec x}_{a}d{\vec x}_{b}, \nonumber
\end{eqnarray}
where vertexes $\Gamma$ are given in (\ref{11})-(\ref{13}) and Kernel ${\widetilde M}^{\rm{4pt\, tree}}_{\psi\,\phi\,|\,\phi\,\psi}$ - in (\ref{47}). 

Integrals over ${\vec x}_{a}$, ${\vec x}_{b}$ that arise in (\ref{52}) after these substitutions can't be taken with elementary formulas (\ref{14}), (\ref{15}). However, as it was shown in \cite{Melon1}, spectral equation for $\Delta_{(\psi\phi)}$ may be obtained from Bethe-Salpeter equation (\ref{52}) in the limit ${\vec x}_{3} \to \infty$. In this limit, according to (\ref{11}), both sides of (\ref{52}) are proportional to $|{\vec x}_{3}|^{- 2\Delta_{(\psi\phi)}}$ that is reduced. Then two boundary integrals in the RHS of (\ref{52}) may be performed using (\ref{15}). 

After these manipulations, with account of (\ref{11})-(\ref{13}), (\ref{15}), (\ref{16}), (\ref{47}) and "extremal" equation (\ref{22}), Bethe-Salpeter equation (\ref{52}) comes to:

\begin{eqnarray}
\label{53}
\frac{B(\Delta_{\psi}, \Delta_{\phi}, \Delta_{(\psi\phi)})}{P_{12}^{\frac{\Delta_{\phi} + \Delta_{\psi} - \Delta_{(\psi\phi)}}{2}}} = g^{* 2}_{I} \,  \frac{B(d - \Delta_{\psi}, d - \Delta_{\phi}, \Delta_{(\psi\phi)}) \, (d - 2\Delta_{\phi}) \, (d - 2\Delta_{\psi})}{P_{12}^{\frac{\Delta_{\phi} + \Delta_{\psi} - \Delta_{(\psi\phi)}}{2}}} \, \cdot \nonumber
\\
\\
\cdot \, C_{\Delta{\phi}} \, C_{\Delta_{\psi}}\, A\left(\Delta_{\psi}, \delta_{1}, \frac{\Delta_{\sigma} + \Delta_{(\psi\phi)}}{2}\right)\, A\left(\Delta_{\psi}, \frac{d + \Delta_{\sigma} - \Delta_{(\psi\phi)}}{2}, \delta_{2}\right),   \nonumber
\end{eqnarray}
where

$$
\delta_{1} = \frac{2d - \Delta_{\phi} - \Delta_{\psi} - \Delta_{(\psi\phi)}}{2}, \qquad \delta_{2} = \frac{d - \Delta_{\psi} - \Delta_{\phi} + \Delta_{(\psi\phi)}}{2}.
$$

Substitution of $C$, $B$, $A$ from (\ref{7}), (\ref{13}), (\ref{16}) and $g^{* 2}_{I}  = 1$ from (\ref{33}) gives the final form of the bootstrap equation (\ref{53}) that determines spectrum of conformal dimension $\Delta_{(\psi\phi)}$  for given conformal dimensions $\Delta_{\phi}$, $\Delta_{\psi}$, $\Delta_{\sigma}$ of primary operators:

\begin{equation}
\label{54}
\frac{\Gamma \left(\frac{\Delta_{(\psi\phi)} - \Delta_{\sigma}}{2}\right) \, \Gamma \left(\frac{d}{2} - \frac{\Delta_{(\psi\phi)} + \Delta_{\sigma}}{2}\right)}{\Gamma \left(\frac{\Delta_{(\psi\phi)} + \Delta_{\sigma}}{2}\right) \, \Gamma \left(\frac{d}{2} - \frac{\Delta_{(\psi\phi)} - \Delta_{\sigma}}{2}\right)} = \frac{\Gamma(\Delta_{\psi}) \, \Gamma\left(\frac{d}{2} - \Delta_{\phi}\right)}{\Gamma(\Delta_{\phi}) \, \Gamma\left(\frac{d}{2} - \Delta_{\psi}\right)}.
\end{equation}

Is not it curious that for $\Delta_{\psi} = d/4$, $\Delta_{\sigma} = d/2$, $\Delta_{\phi} = \Delta_{\psi} + \Delta_{\sigma} = 3d/4$ Eq. (\ref{54}) coincides, up to the factor $-3$, with spectral formula (4.14) $g(h) = 1$ in \cite{Melon1} ($\Delta_{(\psi\phi)}$ here is $h$ in \cite{Melon1})? However in our case $\Delta_{\sigma} = d/2$ is forbidden by spectral equations (\ref{34}) with account of (\ref{32}).

For vertexes $\Gamma_{\Delta_{\sigma}, \Delta_{\phi}, \Delta_{(\sigma\phi)}} ({\vec x}_{1}, {\vec x}_{2}, {\vec x}_{3})$ and $\Gamma_{\Delta_{\psi}, \Delta_{\sigma}, \Delta_{(\psi\sigma)}} ({\vec x}_{1}, {\vec x}_{2}, {\vec x}_{3})$ defined like in (\ref{50}) the Bethe-Salpeter equations analogous to (\ref{52})  with Kernels ${\widetilde M}^{\rm{4pt\, tree}}$ given in (\ref{48}), (\ref{49}) correspondingly are obtained in a similar way:

\begin{eqnarray}
\label{55}
\Gamma_{\Delta_{\sigma}, \Delta_{\phi}, \Delta_{(\sigma\phi)}} ({\vec x}_{1}, {\vec x}_{2}, {\vec x}_{3}) =  \int \, {\widetilde M}^{\rm{4pt\, tree}}_{\Delta_{\sigma},\,\Delta_{\phi}\,|\,\Delta_{\phi},\, \Delta_{\sigma}} ({\vec x}_{1}, {\vec x}_{2}, {\vec x}_{a}, {\vec x}_{b}) \cdot \nonumber
\\
\\
\cdot \, (d - 2\Delta_{\phi}) \, (d - 2\Delta_{\sigma}) \, \Gamma_{d - \Delta_{\phi}, d - \Delta_{\sigma}, \Delta_{(\sigma\phi)}} ({\vec x}_{a}, {\vec x}_{b}, {\vec x}_{3}) \, d{\vec x}_{a}d{\vec x}_{b}, \nonumber
\end{eqnarray}

and

\begin{eqnarray}
\label{56}
\Gamma_{\Delta_{\psi}, \Delta_{\sigma}, \Delta_{(\psi\sigma)}} ({\vec x}_{1}, {\vec x}_{2}, {\vec x}_{3}) =  \int \, {\widetilde M}^{\rm{4pt\, tree}}_{\Delta_{\psi},\,\Delta_{\sigma}\,|\,d - \Delta_{\sigma}, \, d - \Delta_{\psi}} ({\vec x}_{1}, {\vec x}_{2}, {\vec x}_{a}, {\vec x}_{b}) \cdot \nonumber
\\
\\
\cdot \, (d - 2\Delta_{\sigma}) \, (d - 2\Delta_{\psi}) \, \Gamma_{\Delta_{\psi}, \Delta_{\sigma}, \Delta_{(\psi\sigma)}} ({\vec x}_{a}, {\vec x}_{b}, {\vec x}_{3}) \, d{\vec x}_{a}d{\vec x}_{b}, \nonumber
\end{eqnarray}

Finally the following spectral equations for conformal dimensions $\Delta_{(\sigma\phi)}$ and $\Delta_{(\psi\sigma)}$ analogous to Eq. (\ref{54}) are obtained:

\begin{equation}
\label{57}
\frac{\Gamma \left(\frac{\Delta_{(\sigma\phi)} - \Delta_{\psi}}{2}\right) \, \Gamma \left(\frac{d}{2} - \frac{\Delta_{(\sigma\phi)} + \Delta_{\psi}}{2}\right)}{\Gamma \left(\frac{\Delta_{(\sigma\phi)} + \Delta_{\psi}}{2}\right) \, \Gamma \left(\frac{d}{2} - \frac{\Delta_{(\sigma\phi)} - \Delta_{\psi}}{2}\right)} = \frac{\Gamma(\Delta_{\sigma}) \, \Gamma\left(\frac{d}{2} - \Delta_{\phi}\right)}{\Gamma(\Delta_{\phi}) \, \Gamma\left(\frac{d}{2} - \Delta_{\sigma}\right)}.
\end{equation}

and 

\begin{equation}
\label{58}
\frac{\Gamma \left(\frac{\Delta_{\phi} - \Delta_{(\psi\sigma)}}{2}\right) \, \Gamma \left(\frac{\Delta_{(\psi\sigma)} + \Delta_{\phi}}{2} - \frac{d}{2}\right)}{\Gamma \left(d - \frac{\Delta_{(\psi\sigma)} + \Delta_{\phi}}{2}\right) \, \Gamma \left(\frac{d}{2} + \frac{\Delta_{(\psi\sigma)} - \Delta_{\phi}}{2}\right)} = \frac{\Gamma(\Delta_{\psi}) \, \Gamma(\Delta_{\sigma})}{\Gamma\left(\frac{d}{2} - \Delta_{\psi}\right) \, \Gamma\left(\frac{d}{2} - \Delta_{\sigma}\right)} \, \cdot \, \frac{1}{F(\Delta_{\phi})}.
\end{equation}

In general there are infinite number of positive and negative roots of Eq-s (\ref{54}), (\ref{57}), (\ref{58}). We present below some first positive roots of every of these spectral equations in case $d = 4$ for values (\ref{36})-(\ref{38}) of dimensions of primary operators obtained in Sec. 3 from the self-energy bootstrap.

1) For values (\ref{36}), when $\Delta^{(36)}_{\psi} = \Delta^{(36)}_{\sigma} = 4/3$, $\Delta^{(36)}_{\phi} = 4 - \Delta_{\psi} = 8/3$, spectral Eq-s (\ref{54}), (\ref{57}), (\ref{58}) coincide and come to:

\begin{equation}
\label{59}
\frac{\Gamma(\frac{\Delta_{(\psi\phi)}}{2} – \frac{2}{3}) \, \Gamma(\frac{4}{3} – \frac{\Delta_{(\psi\phi)}}{2})}{\Gamma(\frac{\Delta_{(\psi\phi)}}{2} + \frac{2}{3}) \, \Gamma(\frac{8}{3} - \frac{\Delta_{(\psi\phi)}}{2})} = \frac{\Gamma(-\frac{2}{3}) \, \Gamma(\frac{4}{3})}{\Gamma(\frac{2}{3}) \, \Gamma(\frac{8}{3})}.
\end{equation}
Besides two trivial and unphysical roots $\Delta_{(\psi\phi)} = 0$, $\Delta_{(\psi\phi)} = d = 4$ there are following minimal positive roots of (\ref{59}):

\begin{equation}
\label{60}
\Delta^{(36)}_{(\psi\phi)} = \Delta^{(36)}_{(\sigma\phi)} = \Delta^{(36)}_{(\psi\sigma)} = 4,21; \, \, 6,63...
\end{equation}

Omitting analogous simple intermediate steps we present minimal positive roots of spectral Eq-s (\ref{54}), (\ref{57}), (\ref{58}):
 
2) For values (\ref{37}), when $\Delta^{(37)}_{\psi} = \Delta^{(37)}_{\sigma} = 7/5$, $\Delta^{(37)}_{\phi} = 14/5$:    

\begin{equation}
\label{61}
\Delta^{(37)}_{(\psi\phi)} = \Delta^{(37)}_{(\sigma\phi)} = 0,20; \,\, 3,80; \,\, 4,20... \qquad 
\Delta^{(37)}_{(\psi\sigma)} = 4,52; \,\, 6,77...
\end{equation}

3) For values (\ref{38}), when $\Delta^{(38)}_{\psi} = 7/5$; $\Delta^{(38)}_{\sigma} = 6/5$; $\Delta^{(38)}_{\phi} = 13/5$:

\begin{equation}
\label{62}
\Delta^{(38)}_{(\psi\phi)} = 4,52; \,\, 6,77... \qquad \Delta^{(38)}_{(\sigma\phi)} = \Delta^{(37)}_{(\psi\sigma)} = 0,20; \,\, 3,80; \,\, 4,20...
\end{equation}

Although the obtained spectra of correlators (\ref{47})-(\ref{49}) may be of its own interest, the main result of this section is apparently negative: vertex bootstrap equation (second one in (\ref{1})) can't be fulfilled here. The fact is that spectral Eq-s (\ref{54}), (\ref{57}), (\ref{58}) forbid roots $\Delta_{\psi\phi} = \Delta_{\sigma}$ (or $\Delta_{\psi\phi} = d - \Delta_{\sigma}$), $\Delta_{\sigma\phi} = \Delta_{\psi}$ (or $\Delta_{\sigma\phi} = d - \Delta_{\psi}$), $\Delta_{\psi\sigma} = \Delta_{\phi}$ (or $\Delta_{\psi\sigma} = d - \Delta_{\phi}$) correspondingly. That is these spectra do not contain the conformal dimensions of fields responsible for "interaction" in Bethe-Salpeter equations (\ref{52}), (\ref{55}), (\ref{56}).

\section{Conclusion}

\qquad The negative result noted in the end of previous Section is perhaps a signal of incompatibility of conformal bootstrap and exact conformal symmetry. Nature is not conformally invariant. The possible fundamental role of spontaneous breakdown of conformal symmetry was advocated by 't Hooft \cite{Hooft}. 

We shall not discuss here the intriguing problem of spontaneous breakdown of conformal symmetry in context of the AdS/CFT correspondence, but put more simple and in essence a purely mathematical question: to what extent breakdown of conformal symmetry "by hand", for example in a "hard-wall" way in two-branes Randall-Sundrum (RS) model \cite{Randall1}, will change the values of conformal dimensions (\ref{36})-(\ref{38}), (\ref{44}) received in the paper in frames of the "old" conformal bootstrap approach? 

The question is not a simple one since this means that bulk-to-boundary and bulk-to-bulk functions entering self-energy bootstrap equation (\ref{25}) must be taken now on the conformally non-invariant  background - with the loss of crucial conformal simplifications.

If the values (\ref{36})-(\ref{38}), (\ref{44}) of conformal dimensions calculated in the paper in case $d = 4$ are only slightly (supposedly of the order of mass hierarchy $10^{-16}$) modified with the change of background from pure AdS to, say, RS two-branes model, then knowledge of these numbers becomes of great physical importance. 

The point is that in plenty of papers spectra of excitations of different bulk fields in the RS-model are identified with observed masses of particles and resonances - of glueballs, mesons, light and heavy fermions. And even electron neutrino mass scale ($\sim 10^{-3} eV$) and "flavors'" mass hierarchy, which origin is still a mystery, may be predicted when some natural (e.g. "twisted") boundary conditions are imposed on the bulk spinor fields (see e.g. \cite{Neubert}-\cite{Alt3}). These results crucially depend on the values of bulk masses of spinor fields Thus well grounded calculations of bulk masses (that is of conformal dimensions - see e.g. (\ref{6}) for scalars) of different bulk fields may open the way for prediction of plethora of Standard Model's free parameters. And small imaginary parts of conformal dimensions (like in (\ref{44}) for example) may give the width of resonances in frames of this approach. 

In view of these ambitious goals the immediate task for future would be to generalize the considered toy models of scalar fields with their self-energy bootstrap equation (\ref{25}) to more physical models like e.g. bulk Yukawa spinor-scalar model.

\section*{Acknowledgements} Author is grateful to Andrei Barvinsky, Ruslan Metsaev, Arkady Tseytlin and Mikhail Vasiliev for stimulating comments and to participants of the seminar in the Theoretical Physics Department of the P.N. Lebedev Physical Institute for fruitful discussions.

\end{document}